\documentclass[a4paper,12pt]{article}

\usepackage[utf8x]{inputenc}
\usepackage[T1]{fontenc}
\usepackage[colorlinks=true,linkcolor=black,citecolor=black,urlcolor=blue]{hyperref}
\usepackage{geometry}
\usepackage{titlesec}
\titleformat{\section}
{\bfseries\uppercase}{\thesection.}{1em}{}
\titleformat{\subsection}
{\bfseries}{\thesection.\thesubsection.}{1em}{}

\usepackage{graphicx} % used to insert the figure
\usepackage{multirow} % used for the table
\usepackage{cite}
\usepackage{breakurl}
\usepackage{indentfirst}
\usepackage{amsmath, amssymb, amsfonts, bm,stmaryrd}
\usepackage{txfonts}
\usepackage{fourier}
\usepackage{enumitem}
\usepackage{xcolor}
\usepackage{enumitem}

\titlespacing*{\subsection}{0pt}{1.5em}{0.2em}
\titlespacing*{\section}{0pt}{1.5em}{0.2em}

\renewcommand\eqref[1]{Equation~\ref{#1}}

\renewcommand{\thesection}{\arabic{section}}
\renewcommand{\thesubsection}{\arabic{subsection}}

\makeatletter
\renewcommand\@biblabel[1]{#1.}
\makeatother

\newlength{\bibitemsep}
\newlength{\bibparskip}
\let\oldthebibliography\thebibliography
\renewcommand\thebibliography[1]{%
  \oldthebibliography{#1}%
  \setlength{\parskip}{\bibitemsep}%
  \setlength{\itemsep}{\bibparskip}%
}

\newcommand{\YearConf}{2024}

\usepackage{fancyhdr}
  \fancypagestyle{firststyle}
{   \fancyhf{}
}
\begin{document}
\thispagestyle{firststyle}

%\begin{center}
%	\includegraphics[width=2in]{\LogoConf}
%\end{center}
\vskip.5cm

\begin{flushleft}
\fontsize{16}{20}\selectfont\bfseries
%\color{red}(Environmental Sound Classification on An Embedded Hardware Platform)  \\
\color{black}Environmental Sound Classification on An Embedded Hardware Platform
\end{flushleft}
\vskip1cm

\renewcommand\baselinestretch{1}
\begin{flushleft}

Gabriel Bibbó\footnote{g.bibbo@surrey.ac.uk}, Arshdeep Singh\footnote{arshdeep.singh@surrey.ac.uk}, Mark D. Plumbley\footnote{m.plumbley@surrey.ac.uk}\\
%Centre for Vision, Speech and Signal Processing (CVSSP), University of Surrey\\
%Stag Hill, University Campus, Guildford GU2 7XH\\

\vskip.5cm
%Arshdeep Singh\footnote{arshdeep.singh@surrey.ac.uk}\\
%Centre for Vision, Speech and Signal Processing (CVSSP), University of Surrey\\
%Stag Hill, University Campus, Guildford GU2 7XH\\

%\vskip.5cm
%Mark D. Plumbley\footnote{m.plumbley@surrey.ac.uk}\\
Centre for Vision, Speech and Signal Processing (CVSSP), University of Surrey, United Kingdom\\
%Stag Hill, University Campus, Guildford GU2 7XH\\

\end{flushleft}
\textbf{\centerline{ABSTRACT}}\\
\textit{Convolutional neural networks (CNNs) have exhibited state-of-the-art performance in various audio classification tasks. However, their real-time deployment remains a challenge on resource-constrained devices such as embedded systems. In this paper, we analyze how the performance of large-scale pre-trained audio neural networks designed for audio pattern recognition changes when deployed on a hardware such as a Raspberry Pi. We empirically study the role of CPU temperature, microphone quality and audio signal volume on performance. Our experiments reveal that the continuous CPU usage results in an increased temperature that can trigger an automated slowdown mechanism in the Raspberry Pi, impacting inference latency. The quality of a microphone, specifically with affordable devices such as the Google AIY Voice Kit, and audio signal volume, all affect the system performance. In the course of our investigation, we encounter substantial complications linked to library compatibility and the unique processor architecture requirements of the Raspberry Pi, making the process less straightforward compared to conventional computers (PCs). Our observations, while presenting challenges, pave the way for future researchers to develop more compact machine learning models, design heat-dissipative hardware, and select appropriate microphones when AI models are deployed for real-time applications on edge devices.}

\section{INTRODUCTION}
\noindent
Artificial Intelligence (AI) based frameworks  have been successfully employed to solve many real-life applications. For example, AI-based systems  give state-of-the-art performance in problems including  audio classification \cite{kong2020panns} and speech recognition \cite{tu2019speech},  and can be used in various applications such as  assisted living, surveillance, healthcare and activity recognition. However,  the performance of such  well-performing AI-based software systems, when deployed on edge devices,  has not been much explored despite the advancement of compact devices such as embedded systems or micro-controllers. Even though AI has taken impressive strides in academia in the past few years, a gap still exists between theoretical advancements and practical applications. Therefore, this paper aims to analyse the performance of AI models designed for audio classification, when such AI models are deployed onto a compact and a portable hardware device.

We explore the feasibility and challenges of deploying convolutional neural networks (CNNs) on edge devices like the Raspberry Pi \cite{thepihut}, utilizing pre-trained audio neural networks (PANNs) \cite{kong2020panns,ai4sdemo} for audio event classification. Despite advancements since PANNs, we opt for this model due to its simplicity in Linux deployment and effective audio activity recognition.

We assess the performance of PANNs on edge device by capturing  audio using different  microphones, aiming to quantify the effect of microphones on the performance. 
Additionally, we discover difficulties related to temperature and volume of the audio signal on the performance of the PANNs models on the edge device. We also encounter challenges in installing the PANNs code on a Raspberry Pi with the Raspbian operating system. Unlike installations on conventional computers (PCs), we require specific libraries and specific compilations for the Raspberry Pi's processor architecture. We also find that running AI algorithms can significantly increase the  temperature of the CPU in the Raspberry Pi after few minutes of operation. To prevent overheating, the Raspberry Pi automatically reduces the CPU clock speed, resulting in an increased inference latency. 
To summarise our contributions, we aim to answer the following questions:

\begin{itemize}
    \item What are the challenges in deploying AI models on edge devices, particularly the Raspberry Pi?
    \item How does device temperature  affect the inference latency?
    \item How does microphone quality and volume impact PANNs performance in recognising audio events on edge device?
\end{itemize}

We hope our work shines a light on existing and emerging challenges for the detection and classification of acoustic scenes and events (DCASE) research  community, promoting discussion on deploying advanced algorithms in real-world settings. We envision our work as a practical guide for deploying similar systems on embedded devices, aiming to ease the path for their broader adoption.

The rest of the paper is organised as follows. Section \ref{sec: background} introduces some background  on assisted living applications and edge devices. Section \ref{sec: system requrements} presents various components used in developing the hardware-based demonstration and the objectives to perform exterminations.  Next, the experimental setup is explained in Section \ref{sec: experimental}. Section \ref{sec: results} presents experimental analysis. Finally, Section \ref{sec: discuss} discusses the issues encountered in development and  Section \ref{sec: Conclusion} concludes the paper. 

\section{Related work}
\label{sec: background}
\noindent
\subsection{Everyday ambient assisted living using audio}
\label{sec:AAL}
\noindent
Frameworks for the detection of acoustic events have shown significant advances in  applications, including the monitoring of elderly or dependent individuals \cite{wang2019privacy, vafeiadis2020audio,rashidi2012survey} and surveillance and security applications \cite{schwartz2012chicago,chaudhary2018identification, alias2019review}. Recognizing certain interior sounds has proven beneficial in multiple monitoring contexts, particularly those related to human behaviour \cite{wang2019privacy}. This becomes practically important when integrated into smart homes for ambient assisted living \cite{vafeiadis2020audio}, to aid the elderly or individuals with disabilities in their daily lives, improving their overall quality of life. Furthermore, technological progress, particularly the advent of the Internet of Things (IoT), has enabled these DCASE algorithms to be deployed in real-time wireless acoustic sensor networks \cite{alias2019review}. This has enabled innovative projects for the support of the elderly in their homes, such as CIRDO \cite{bouakaz2014cirdo} and homeSound \cite{alsina2017homesound}. These initiatives focus on home safety, enabling medical staff to remotely monitor the status of dependent individuals using a decentralised intelligence architecture. Moreover, detecting sounds in outdoor environments have several applications including outdoor surveillance \cite{chaudhary2018identification}, traffic noise mapping \cite{socoro2015development} and  soundscape modelling \cite{jeon2015classification}.

\subsection{Edge devices and their challenges}
\label{sec:tinyml}
\noindent
While developing real-world applications, it is important to understand the landscape of available hardware options for deployment, in order to optimize cost, hardware size, and software used in the development stage. One of the common solutions is to use edge devices  which  may offer a promising platform for DCASE algorithms in real-time applications due to their resource-constrained, small size, low energy consumption and affordability \cite{gruenstein2017cascade, koizumi2019toyadmos, chowdhery2019visual}. However, there are several challenges to the deployment of resource-hungry AI models such as CNNs on resource-constrained edge devices. These challenges include the limited memory and computation capacity of edge devices \cite{banbury2021mlperf}, hardware heterogeneity, software fragmentation, and the necessity of optimizing and coordinating across different layers (hardware, software, algorithms, etc.) of the development and deployment stack. Additionally, the operation of edge devices on small, low-capacity batteries imposes further restrictions \cite{siekkinen2012low}, such as limited battery life, the need for low-power modes and reduced duty cycles that can affect performance and latency, constraints on transmission power and communication range, and potential tradeoffs in functionality to prolong battery life. Therefore, it is important to understand the limitations of edge devices used to deploy AI-based software in real-life.

\section{System requirements and functionality}
\label{sec: system requrements}
\noindent
\subsection{Hardware used for experimentation}
\noindent
For our design, we require the ability to handle real-time processing tasks, such as audio monitoring, and allow complex digital signal processing to be performed on the device itself, minimizing the need to transmit large amounts of audio data for centralized computation. With these requirements in mind, at the core of our system is the Raspberry Pi 4 \cite{thepihut} processor, which is a 64-bit Quad-core Cortex-A72 (ARM v8) 1.8 GHz system on chip, with 4 GB RAM, 64 GB flash storage, USB I/O, and Wi-Fi connectivity. This hardware platform, running a Linux-based operating system, provides a range of AI libraries and community support. The Raspberry Pi 4 is easy to use and requires minimal software code adaptation.

\begin{figure}
    \centering
    \includegraphics[scale = 0.8]{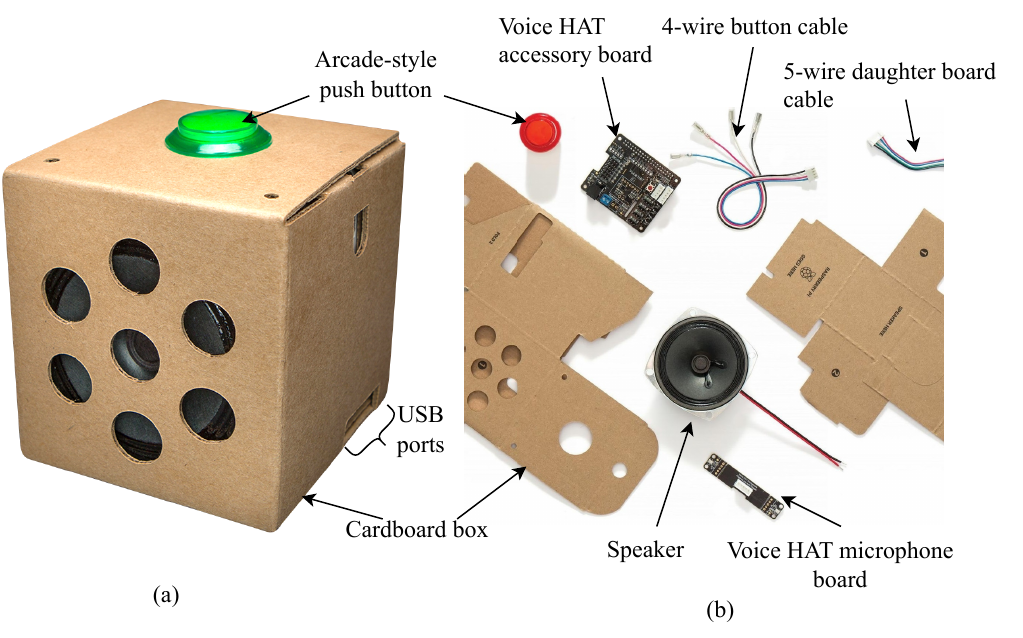}
    \vspace{-0.6cm}
    \caption{Google AIY Voice Kit (a) assembled, (b) components}
    \label{fig: Google AIY box}
\end{figure}

During our quest for suitable hardware, we identified the Google AIY Voice Kit \cite{google2023}, an AI project designed for voice detection, that is closely aligned with our requirements. The kit is designed to encourage hands-on learning about artificial intelligence and programming in an accessible manner, and was originally created to facilitate the prototyping of voice AI applications on the Raspberry Pi. We adapt the AIY Voice Kit hardware and software package to build an audio tagging framework. The AIY Kit and its various components are shown in Figure  \ref{fig: Google AIY box} (a). The components, as shown in Figure  \ref{fig: Google AIY box} (b), include a Voice HAT (Hardware Attached on Top), an ICS-43432 stereo microphone optimized for speech detection, cables and a speaker, all housed in an environmentally friendly cardboard box \cite{google2023}.

\subsection{Google AIY Voice Kit software}
\noindent
The base operating system in the Google AIY Voice Kit is Raspbian version 5.1.0.17-v7l+, which comes preinstalled with a default image. Raspbian is an open source, linux-based operating system specifically designed for Raspberry Pi hardware. It consumes minimal computational resources and allows control over various functionalities that the kit offers, from integration with button control libraries, both for pressed action and LED light speaker and microphone, as well as text-to-speech capabilities.

\subsection{Audio tagging software}
\noindent
For recognising audio events in the surroundings, we use PANNs-based Aritificial Intelligence for Sound (AI4S) demonstration software \cite{ai4sdemo}, a software suite that includes a graphical interface for visualising and assigning confidence/probability values corresponding to detected audio events. The AI model in the software includes a pre-trained convolutional neural network (PANNs) \cite{kong2020panns} designed  for audio pattern recognition. We use the CNN9 model from PANNs that  achieves a mean Average Precision (mAP) of 0.37 and has 4.96 million parameters. This PANNs model has been trained on AudioSet \cite{audioset}, a large-scale dataset of approximately 2 million audio clips with 527 distinct audio event classes. PANNs takes as input the log-mel spectrogram of sound signals. We apply certain usability modifications to the code to better align it with our project's specific needs, leaving the core audio event detection algorithm unaltered. The resulting code graphical user interface (GUI) shows the top few predicted audio event classes with their confidence values.

\subsection{Functionality of audio tagging system on hardware}
\noindent
The audio tagging system hardware is controlled  by a push button, without the need of a keyboard, mouse, or screen. Upon booting, it automatically connects to the network and plays a welcome audio message: \textit{This is the AI4S demo. Press the button to start recording}. When the push button (ON/OFF) is pressed, an LED lights up and sound recording begins in a buffer. Then, the recording signal is processed by the audio tagging software explained previously to generate and notify the top few predicted audio classes corresponding to the recorded audio signal in the buffer.

The audio tagging system on hardware hosts a webpage accessible from any device on the same local network  for debugging and remote access. It provides a real-time view of both the CPU temperature, as reported by the Raspberry Pi, and a log of detected audio events, with a ranking of the most frequently appearing sounds. Additionally, the system  has the capability to send email alerts when specific sounds, such as a fire alarm, water tap etc. are detected. This feature offers an additional layer for real-time safety, and can be personalised depending on the used case.

\subsection{What  do we want to measure?}
\noindent
With the audio tagging system on hardware or software, we aim to measure or compare the following:\\

\begin{itemize}[noitemsep]
\item
\textbf{Hardware v/s software performance:} The performance of the Raspberry Pi based audio tagging system and the software-based audio tagging system which runs only on a PC;\\
\item
\textbf {Performance v/s volume of audio signal:} Effect of audio signal volume on performance;\\
\item
\textbf{Performance v/s microphone quality:} Effect of different microphones on the performance;\\
\item
\textbf{Temperature v/s latency:} The temperature and latency in the Raspberry Pi based audio tagging system.
\end{itemize}

\begin{figure}[t]
  \centering
  \includegraphics[scale = 0.9]{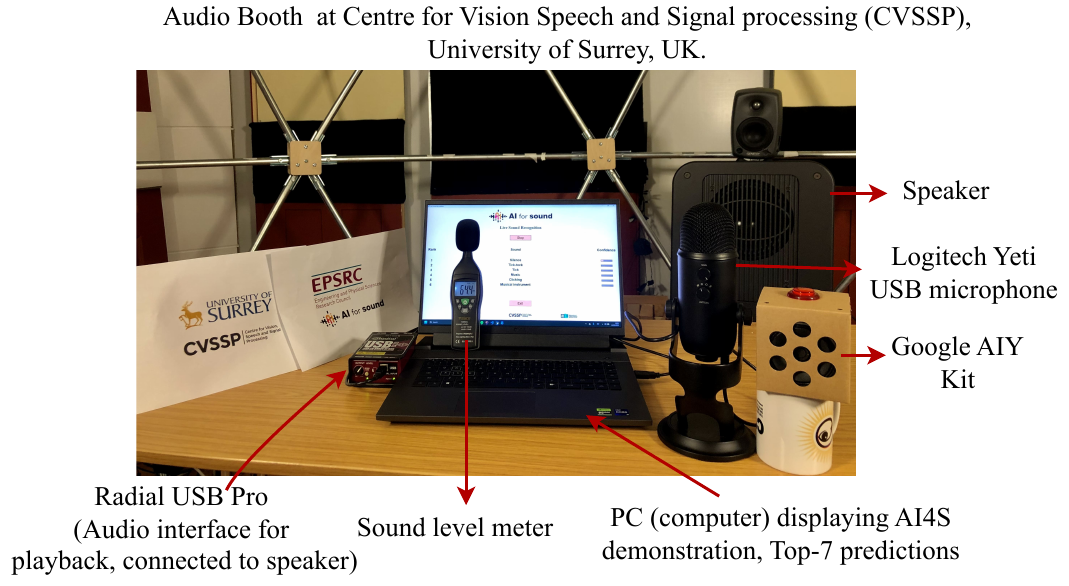}
  \vspace{-0.85cm}
  \caption{Experimental setup}
  \label{fig:audio lab}
\end{figure}

\section{Experimental Setup}
\label{sec: experimental} 

\subsection{Dataset preparation \& experimental environment}  
\noindent
For experimentation, we selected five audio categories which are closely aligned with the objectives of AI4S project, on wellbeing and sound at home. The five audio categories are ``Speech'', ``Baby cry'', ``Water'', ``Fire alarm'' and ``Music'' from  AudioSet \cite{audioset}. 

Next, for each audio class, we created a continuous two-minute-long audio clip by concatenating 10 audio examples from each class, sourced from AudioSet. For example, we selected  10 examples of alarm sounds and concatenated them to create 2 minutes of alarm sounds. After obtaining the 2-minute concatenated audio file, we removed silence and normalized the volume. In the end, we have five different audio recordings each 2 minutes in length, one for each class.

\begin{itemize}[noitemsep]
\item
\textbf{Varying playback volume}: We prepare 50, 60, and 70dB volume levels for each audio recording to measure the impact of volume on system performance. These volumes were determined using a N05CC sound level meter \cite{precisiongold2023} in dBA mode with the drivers at a distance of one meter from the microphone. The dBA mode is used for general sound level measurements. We choose these three different volume levels based on the potential range of sound events encountered in home or office environments \cite{mehta2012noise}. 
  
\item
\textbf{Acoustic Environment and audio playback}: We perform experiments in the Audio Booth in the Centre for Vision, Speech, and Signal Processing (CVSSP) at the University of Surrey, UK. The experimental setup is shown on Figure \ref{fig:audio lab}. The Audio Booth is an isolated room without interference from outside noise, and has a base noise of 34 dB sound pressure level inside the room. For audio playback, we simulate a wide-frequency sound source located one meter away from the microphones by utilizing a Genelec 8020B Monitoring System loudspeaker and a 7060A Active Subwoofer, one over the other, as can be seen in Figure \ref{fig:audio lab}.
\end{itemize}

\subsection{ Systems used to compare} 
\label{sec: system for comparison}
\noindent
We use three distinct scenarios to compare the performance of the PANNs model performance to classify the five audio recordings  on hardware and software:

\begin{itemize}[noitemsep]
\item
\textbf{Google AIY Mic + Rasp\_Pi}: We utilize the assembled device, equipped with the low-cost AIY ICS-43432 \cite{InvenSense2023} microphone, and a Raspberry Pi running the audio tagging software. For audio recording playback at different volumes, the audio events are played through loudspeakers.
\item
\textbf{Yeti Mic + PC}: We utilize a  Logitech Yeti USB microphone to capture the audio events and the audio tagging software running on a dedicated PC. Similar to the scenario (1) above, the audio events at different volumes are played through loudspeakers.
\item
\textbf{PC with recorded audio}: We evaluate the performance of the audio tagging software directly on the computer (PC) without involving any microphone and playback. The audio recordings at different volume levels are stored in the computer and used directly to compute performance.
\end{itemize}

\subsection{Performance metrics} 
\noindent
To evaluate the performance of the audio tagging software on the Raspberry Pi and the PC, we use a sliding window approach to analyze the audio recordings. The software processes the audio using a 2-second sliding window, making predictions for each window. Based on these predictions, we calculate two metrics for each distinct sound event class:

\textbf{Mean Confidence and Standard Deviation}: For each sliding window, the software provides confidence values for each sound event class, representing the model's belief that a specific sound event is present within that window. We calculate the mean and standard deviation of these confidence values across all windows in the audio recording for each specific sound event class.

\textbf{Percentage of Occurrence}: We consider a specific sound event to have occurred within a window if it appears in the Top-7 predictions for that window; otherwise, it is considered absent. We then calculate the percentage of occurrence for each sound event class by counting the total number of windows in which the event was detected and dividing it by the total number of windows analyzed in the entire audio recording. This percentage represents the proportion of windows in which a specific sound event was detected.

\section{Experimental Analysis}
\label{sec: results}
\noindent
Table \ref{table:results} shows the mean confidence scores and percentage of occurrence from different systems explained in Section 4.\ref{sec: system for comparison} at varying audio signal playback volumes. Our experiments led to the following observations:

\begin{table}[t]
\centering
\caption{Mean ($ \mu $) and standard deviation ($ \sigma $) of confidence values, and percentage of occurrence as one of the Top-7 predictions using different systems for various sound events at varying volumes.}
\Large
\resizebox{.8\textwidth}{!}{\begin{tabular}{l c c c c c c c}
\hline

\multirow{2}{*}{\shortstack[l]{\textbf{Sound}\\ \textbf{Class}}} & \multirow{2}{*}{\shortstack[l]{\textbf{Vol}\\ \textbf{(dB)}}} & \multicolumn{2}{c}{\textbf{Google AIY Mic + Rasp\_Pi}} & \multicolumn{2}{c}{\textbf{Yeti Mic + PC}} & \multicolumn{2}{c}{\textbf{PC with recorded audio}} \\
\cline{3-8}
 & & $ \mu\pm\sigma $ & Top-7  & $ \mu\pm\sigma $ & Top-7  & $ \mu\pm\sigma $ & Top-7 \\
\hline
\multirow{3}{*}{Speech} & 50 & 0.69$\pm$0.17 & 100\% & 0.51$\pm$0.21 & 99.7\% &0.75$\pm$0.12  &  100\% \\
 & 60 & 0.72$\pm$0.15 & 100\% & 0.52$\pm$0.21 & 100\% &0.77$\pm$0.12  & 100\% \\
 & 70 & 0.73$\pm$0.14 & 100\% & 0.54$\pm$0.23 & 99.7\% &0.78$\pm$0.12  & 100\% \\
\hline
\multirow{3}{*}{\shortstack[l]{Baby\\ cry}} & 50 & 0.15$\pm$0.15 & 70.8\% & 0.20$\pm$0.16 & 62.7\% & 0.30$\pm$0.24 & 93\% \\
 & 60 & 0.22$\pm$0.19 & 82.8\% & 0.19$\pm$0.15 & 64.5\% & 0.34$\pm$0.25  & 95\%  \\
 & 70 & 0.32$\pm$0.25 & 84.0\% & 0.23$\pm$0.18 & 66.4\% & 0.37$\pm$0.26 & 95\%  \\
\hline
\multirow{3}{*}{\shortstack[l]{Water}} & 50 & 0.13$\pm$0.10 & 46.5\% & 0.21$\pm$0.16 & 81.9\% &  0.28$\pm$0.21 & 85\%  \\
 & 60 & 0.16$\pm$0.15 & 60.1\% & 0.33$\pm$0.21 & 85.3\% & 0.28$\pm$0.22 & 83\%  \\
 & 70 & 0.22$\pm$0.16 & 58.0\% & 0.32$\pm$0.20 & 80.2\% &  0.17$\pm$0.20 & 62\%  \\
\hline
\multirow{3}{*}{\shortstack[l]{Fire\\ alarm}} & 50 & 0.11$\pm$0.08 & 70.4\% & 0.42$\pm$0.21 & 95.9\% & 0.32$\pm$0.25 & 92\%  \\
 & 60 & 0.15$\pm$0.09 & 68.9\% & 0.55$\pm$0.21 & 97.5\% & 0.39$\pm$0.27  & 92\%  \\
 & 70 & 0.22$\pm$0.13 & 71.2\% & 0.68$\pm$0.21 & 98.5\% & 0.45$\pm$0.27 &  95\%\\
\hline
\multirow{3}{*}{\shortstack[l]{Music}} & 50 & 0.49$\pm$0.26 & 93.7\% & 0.56$\pm$0.23 & 98.0\% & 0.67$\pm$0.23 & 100\% \\
 & 60 & 0.55$\pm$0.26 & 96.9\% & 0.64$\pm$0.22 & 99.1\% & 0.72$\pm$0.24 & 100\% \\
 & 70 & 0.61$\pm$0.24 & 97.8\% & 0.71$\pm$0.21 & 99.5\% & 0.76$\pm$0.25 & 100\%  \\
\hline
\end{tabular}}
\label{table:results}
\end{table}

\subsection{Performance differences between the embedded system and the software system}
\noindent
The PC with recorded audio (scenario (3)) shows  an improvement of at least 5 percentage points in confidence and a minimum of 3 percentage points improvement in Top-7 predictions compared to both the Raspberry Pi-based system (scenario (1)) and the Yeti microphone-based PC system  (scenario (2)) across all audio classes and audio volume levels except for  ``Water'' class at 70dB.

\subsection{Impact of playback volume}
\noindent
Generally, the  performance obtained using different systems increase with an increase in volume of the audio signals for all sound classes except for ``Water'' sounds. We perceptually find that increasing the volume for ``Water'' makes the audio recording more noisy. We also confirm  using Audacity audio software \cite{audacity2017audacity} that the ``Water'' sound at 70 dB saturates resulting in audio clipping that may cause the performance to degrade compared to the lower volume sounds. 

\subsection{Impact of different microphones}
\noindent
We find that the AIY microphone outperforms the Logitech Yeti in the detection of ``Baby cry'', and ``Speech'' sounds for majority of the sound volume levels. It is worth mentioning that the AIY microphone is specifically designed for spoken voice detection \cite{InvenSense2023}, which may explain why it is able to detect speech and baby cry sounds better than that of the Logitech Mic.  On the other hand, the Logitech Yeti microphone performs better than the Google AIY microphone for other sounds.

\subsection{Performance of Audio tagging system}
\noindent 
The audio tagging system using the PANNs model on the Raspberry Pi or PC shows significantly better performance in detecting ``Music'' and ``Speech'' sounds compared to other sounds. This might be due to the pre-trained PANNs model which is trained using more  speech and music audio class examples compared to that of the other sounds \cite{audioset}. 

\section{Discussion: Development Issues}
\label{sec: discuss}
\noindent 

\subsection{Installation in ARM-based CPUs} 
\noindent 
Deploying the PANNs-code on a Raspberry Pi with the Raspbian operating system is not a straightforward process compared to that on a conventional computer. On the Raspberry Pi, specific library versions need to be used, and in some cases, they need to be compiled specifically for the ARM architecture of the Raspberry Pi. We provide a GitHub repository \cite{pisoundsensing} giving a detailed step-by-step guide on how to install the demonstration correctly on Raspberry Pi to promote the easier usability and easier deployment of the hardware-based AI frameworks.

\subsection{Temperature and latency of Raspberry Pi} 
\noindent
We find that the temperature of the Raspberry Pi's CPU increases when we run the AI  algorithms on it. To prevent overheating, the Raspberry Pi has an automatic CPU clock control mechanism that slows down the processing on the Raspberry Pi whenever the temperature gets too high, resulting in increased inference latency.

\begin{figure}[t]
  \centering
  \includegraphics[width=0.9\textwidth]{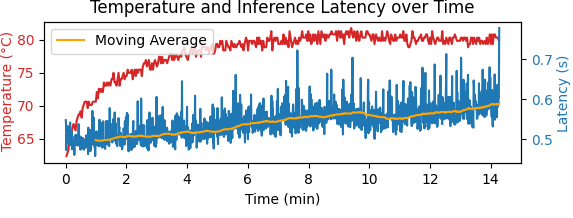}
  \vspace{-0.75cm}
  \caption{Temperature and inference latency over time when audio tagging system is  running on Raspberry Pi.}
  \label{fig:tempvslatency}
\end{figure}

Figure \ref{fig:tempvslatency} shows the  the temperature and latency in the Raspberry Pi when the PANNs model runs for few minutes. We observe that the there is a rise in temperature of more than 25$^{\circ}$C after 14 minutes  of continuous operation. After 8 minutes  of running the PANNs model, the CPU temperature stabilises around 79$^{\circ}$C.  Meanwhile, as the temperature increases, the average inference latency also increases from approximately 0.5s to 0.6s. To mitigate this effect, it is important to incorporate heat sinks or a ventilation system for the Raspberry Pi hardware container to provide sufficient cooling.

\section{Conclusions and future work}
\label{sec: Conclusion}  
\noindent
In this paper, we presented a deployment of a PANNs-based audio tagging framework on a Raspberry Pi hardware system. Our findings suggest that the performance degrades when AI models are deployed on a Raspberry Pi compared to software-based (PC) performance. We also find that the selection of an appropriate microphone is an important factor in recognizing audio events, and the volume levels of the audio events may limit the performance of the AI models.

We observe that effective device temperature management is important to maintain optimal performance and reduce inference latency. When exploring heat control methods like fans, it is important to consider the negative impact they may have on the audio estimates made by the device. Fans generate noise, known as egonoise, which can reduce the quality of the audio captured by the microphone. Attaining high accuracy is a balancing act involving microphone type, audio signal volume, and the chosen embedded system.

In the future, we would like to explore more experiments covering a wider variety of audio event classes to understand the impact of real-world versus recorded sound sources on performance. Analyzing different neural network architectures beyond PANNs would also be valuable. Moreover, we are interested in identifying the range of audio volumes suitable for optimal operation of these edge devices, determining the minimum and maximum levels within which the system can operate reliably, and designing appropriate heat control measures to increase the efficiency of hardware-based edge devices.

Furthermore, future studies could investigate techniques to optimize models for edge devices, such as pruning or knowledge distillation, to reduce computations and enable faster inference. Measuring the energy consumption of the Raspberry Pi during inference would provide insights into performance-power trade-offs. Considering a broader range of edge devices beyond the Raspberry Pi, such as microcontrollers and FPGAs, would provide a more comprehensive understanding of the challenges and opportunities in deploying AI models on resource-constrained devices.

In conclusion, our work serves as a starting point for investigation into the deployment of audio tagging models on edge devices. Despite the limitations, we believe our findings will help guide the next directions of research in this area. By addressing the identified challenges and exploring the suggested future directions, we can strengthen the applicability and impact of audio tagging models in real-world edge computing scenarios.

\section*{Acknowledgements}
\noindent
This work was supported by Engineering and Physical Sciences Research Council (EPSRC) Grant EP/T019751/1 ``AI for Sound (AI4S)''. For the purpose of open access, the authors have applied a Creative Commons Attribution (CC BY) licence to any Author Accepted Manuscript version arising. %We would like to thank Dr Tim Brookes, from the Institute of Sound Recording (IoSR) at the University of Surrey, for his invaluable advice and for providing the volume meter Precision Gold IEC 651. 

\bibliographystyle{unsrt}
\bibliography{main} 

\end{document}